# InAs Nanowire-Based Twin Electrical Sensors Enabling Simultaneous Gas Detection Measurements: Nanodevice Engineering, Testing and Signal Fluctuation Analysis

Camilla Baratto,*,† Egit Musaev,‡,†,# Valeria Demontis,¶,§,# Stefano Luin,§ Valentina Zannier,§ Lucia Sorba,§ Guido Faglia,‡,† Luigi Rovati,| and Francesco Rossella*,⊥

†CNR-INO PRISM Lab, Via Branze 45, Brescia, Italy
‡Department of Information Engineering,Via Branze 38, Brescia, Italy
¶Department of Physics, University of Cagliari, S.P. Monserrato-Sestu, Monserrato, 09042, Italy.
§NEST Laboratory, Scuola Normale Superiore and Institute of Nanoscience – CNR, piazza San Silvestro 12, Pisa 56127, Italy.
|Department of Engineering "Enzo Ferrari" , University of Modena and Reggio Emilia, Via Vivarelli, 10, building 26, I-41125 Modena, Italy.
⊥Dipartimento di Scienze Fisiche, Informatiche e Matematiche, Università di Modena e Reggio Emilia, Via Campi 213/a, I-41125 Modena, Italy.
#These authors contributed equally

E-mail: camilla.baratto@cnr.it; francesco.rossella@unimore.it

**Abstract**

Epitaxially grown InAs NWs are relevant for electrical sensing applications due to Fermi level pinning at NW surface, thus very sensitive to surrounding environment. While a single NW growth batch consists of millions virtually identical replicas of the same NW, real samples display subtle differences in NW size, shape, structure which may affect the detection performance. Here, electrical gas detection is investigated the in two NW-based nominally identical or twin devices fabricated starting from the same NW growth batch. Two individual wurtzite InAs NWs are placed onto a fabrication substrate at 2 $\mu$m distance with 90 degrees relative orientation, each NW is electri- cally contacted, and the nanodevices are exposed to humidity and $NO_2$ flux diluted in synthetic air. Electrical signal versus time is measured simultaneously in each nanode- vice, upon different gases and concentrations. Observed detection limit is 2 ppm for $NO_2$, 20% for relative humidity. Correlation analysis method is exploited by calculat- ing auto- and cross-correlation functions for the experimental signal pairs, indicating lack of cross-correlation in the signal noise of the two nanodevices, suggesting that signal differences could be ascribed mainly to nonidealities of fabrication protocol and nanoscopic differences in the two nanostructures, rather than different environmental conditions.

# Introduction

Sensors have become essential devices in many modern technologies, driving the quest for low-cost, easily addressable solutions for several applications, especially environmental gas detection.[1] The advancement of nanotechnology and nanofabrication techniques has en- abled the development of novel and advanced nanodevices for sensing applications, based on nanostructures such as semiconductor nanowires (NWs).[2] Multiple types of sensors based on individual semiconductor NWs or NW arrays[3–5] using both optical and electrical signal transduction principles have been largely demonstrated.[6–9] Among these, conductometric sensors base their operation on current modulations induced by surface interactions between gases and the electrical conductor, and stand out as cost-effective and reliable options. In a gaseous environment, sensing based on semiconductors is promoted by either charge transfer from/to adsorbed molecules or by gas-induced alteration of the height of Schottky barriers generated at the metal-semiconductor contacts[10].[11] NWs are especially suitable for appli- cations in gas sensing because they maximize the surface-to-volume ratio, allowing for high sensitivity, and display well defined crystalline facets, ensuring stability (e.g. over time). In fact, effective sensors exploiting direct electrical readout have been engineered by using semi- conductor NWs[12][13][14],[15] and in general these systems are considered having a huge potential for next-generation chemical sensors.

On the one hand, a family of experimental studies focuses on the performance of ensem- bles of NWs contacted by two global electrodes[16].[17] By using this approach, the estimation of the electrical properties of NWs, and consequently the device performances, are mediated among all the nanostructures composing the array, although the individual nano-objects can differ one from each other. The macroscopic contacts connect, ideally in parallel, thousands of NWs, although the global electrical properties are affected by the potential barrier between two or more different NWs touching each other,[18] like in nanostructured films, where grain boundary rules the conduction. On the other hand, single NW-based sensors are regarded as greatly promising for specific applications. For instance, in the case of metal oxides based gas sensors such as $SnO_2$ and ZnO, fundamental works on the behavior of the single NW have enabled a quite deep comprehension of the sensing mechanism[19][20].[21] However, metal oxides based sensors have the drawbacks of high temperature operation, [22] and resort to UV induced adsorption/desorption phenomena for their room temperature operation.[23]

Epitaxially grown InAs NWs are regarded as a very promising nanomaterial platform for the development of gas sensors operating at room temperature. In these nanostructures, the occurrence of Fermi level pinning at the surface promotes the presence of high-density surface states and the onset of an accumulation layer, which can be extremely useful for sens- ing the environment surrounding the NW surface, particularly for detecting the presence of gases that can be absorbed as charged molecules.[24] InAs NWs in array configurations have been exploited for ethanol[25] as well as $NO_2$ [26] detection at very low concentration, diluted in $N_2$ atmosphere. Regarding individual InAs NW-based sensors, it has been reported their suitability for humidity and organic vapor detection in an inert atmosphere (either Nitro- gen or Helium)[24][27],[28] and different device configurations have been investigated.[6] Naively, all the NWs isolated from the same growth batch can be regarded as identical replicas of the same ideal structure. However, in real samples, differences between the NWs are un- avoidable, and depending on the targeted applications, such differences may play a minor role or may represent a major drawback. For instance, NW-to-NW property variations may result in small changes of the electrical response of devices fabricated starting from different NWs. Moreover, the same nanodevice fabrication protocol applied to identical NWs may lead to slightly different devices. Therefore, an open issue regarding any type of electrical sensors developed starting from a single nano-objects, such as single InAs NWs, is the depen- dence of the device response upon the characteristics of the specific nanostructure, namely crystalline purity, morphology and dimensions. Moreover, also the specific device features, namely nanostructure orientation with respect to electrodes, as well as electrode materials, dimensions and thickness, may affect the detection performance. Indeed, all the potential differences between two ideally identical nanoscale sensors represent in turn potential sources of differences in the response signals of the two devices, and may induce artificial noise in the measurements.

In this work, we report the design, realization and experimental measurement, at room temperature, of two InAs NW-based nominally identical (twin) devices operating as electri- cal sensors of gases. The two NW-based sensors were fabricated starting from nanostructures with diameter difference not exceeding 10 nm isolated from the same growth batch and the architecture was designed in order to ensure that the two nanowires were exposed to the gas flow in the same conditions. Specifically, the two devices were placed at approximately 2 $\mu$m distance with 90 degrees relative orientation. A custom readout interface was developed to enable the simultaneous measurement of the two devices during the sensing experiments. Based on the experimental outcomes, we carried out the signal correlation analysis resorting to the calculation of auto-correlation and cross-correlation functions of the measured electri- cal current signal noise (estimated as residuals from fits), upon device exposure to different gases with different concentrations. We found an overall lack of cross-correlation, on the time scale of seconds, in the electrical noise response of the two nanodevices measured in the same conditions. The different fluctuations in the responses of the two detectors are not ascribable to different environments surrounding locally the surfaces of the two NWs, but instead to unidealities of the fabrication protocol combined to slight (nanoscopic) differences in the individual nanostructures.

## Results and discussion

### Nanodevices architecture

For the present work, InAs nanowires with diameter $70\pm5$ nm are isolated from the same growth batch and used for the fabrication of pairs of nanowire-based nominally identical electrical sensors, hereafter referred to as twin sensors or twin devices. NWs are transferred onto a $Si^{++}/SiO_2$ fabrication substrate, and devices are realized starting from NW pairs placed approximately 2 $\mu$m apart, oriented at 90 degrees relative to each other. Each NW of the pair is equipped with two nominally identical electrical contacts, realized by evap- orating two metallic electrodes ($610\pm10$ nm wide) on both ends of the NW, at a distance of $1400\pm5$nm. In our experiments, a gas flux of synthetic air containing the gas analyte (humidity and $NO_2$) flows onto the NW twin devices, while electrical current is measured simultaneously in both nanodevices as function of time. A custom readout interface enables the simultaneous measurement of the two NWs.

Figure 1 shows a pictorial view of the pair of InAs NW-based devices used in this work, as well as the false-colored scanning electron micrograph of one of the fabricated twin sensors (top-view), together with an example of the measurement output consisting in two resistance-versus-time curves acquired simultaneously in the two NWs.

Notably, while the difference in diameter between the two nanostructures is likely not exceeding 10 nm, this might yield to a fairly significant difference in the surface of the nanostructure exposed to the gas, corresponding to a surface/volume ratio variable in the range from 0.067 (for 60 nm diameter nanowire) to 0.057 (for 70 nm diameter nanowire). Developed sensors might therefore be very sensitive to the morphological parameters of the nanostructures and the metallized pattern, yielding to slight differences in the electric response from device to device.

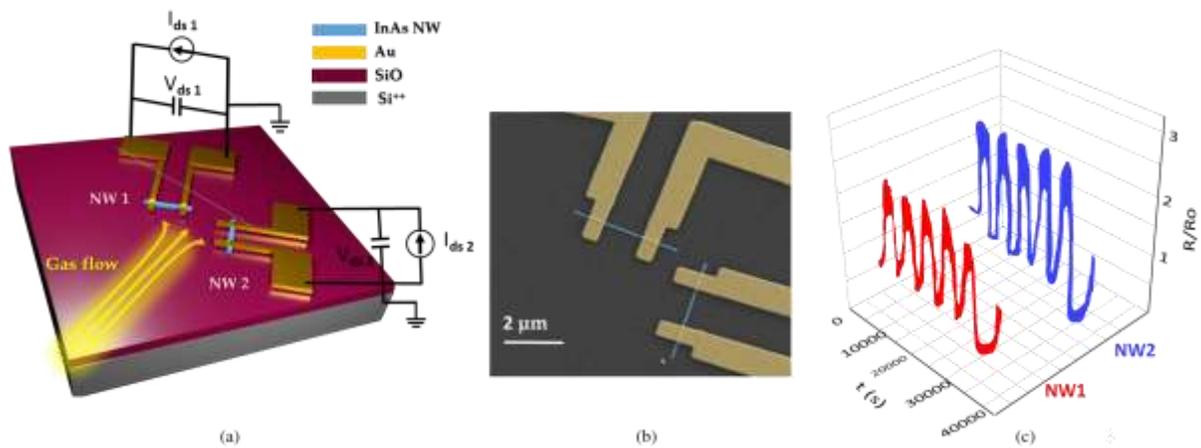

Figure 1: a) Schematic of the two InAs NW-based devices used in this work, b) scanning electron micrograph of one of the fabricated nanowire twin device, c) schematics of the measurement output.

## Simultaneous gas detection measurements

### RH detection

We evaluated the twin NW devices response to variations of relative humidity at levels ranging from 10% to 70% RH in 10% increment at room temperature. Each testing cycle consisted of a 1-hour exposure period during which the sensors were subjected to an input air flow with specific and constant humidity concentration, followed by a 1-hour recovery period during which the samples were allowed to recover to their baseline resistance in dry air. The baseline resistance of the NWs typically ranged between 12 kΩ and 25 kΩ. The experimental outcomes are presented in Figure 2, which summarizes the performance of the NW twin devices as normalized resistance plotted against time, evidencing remarkable response to humidity exposure.

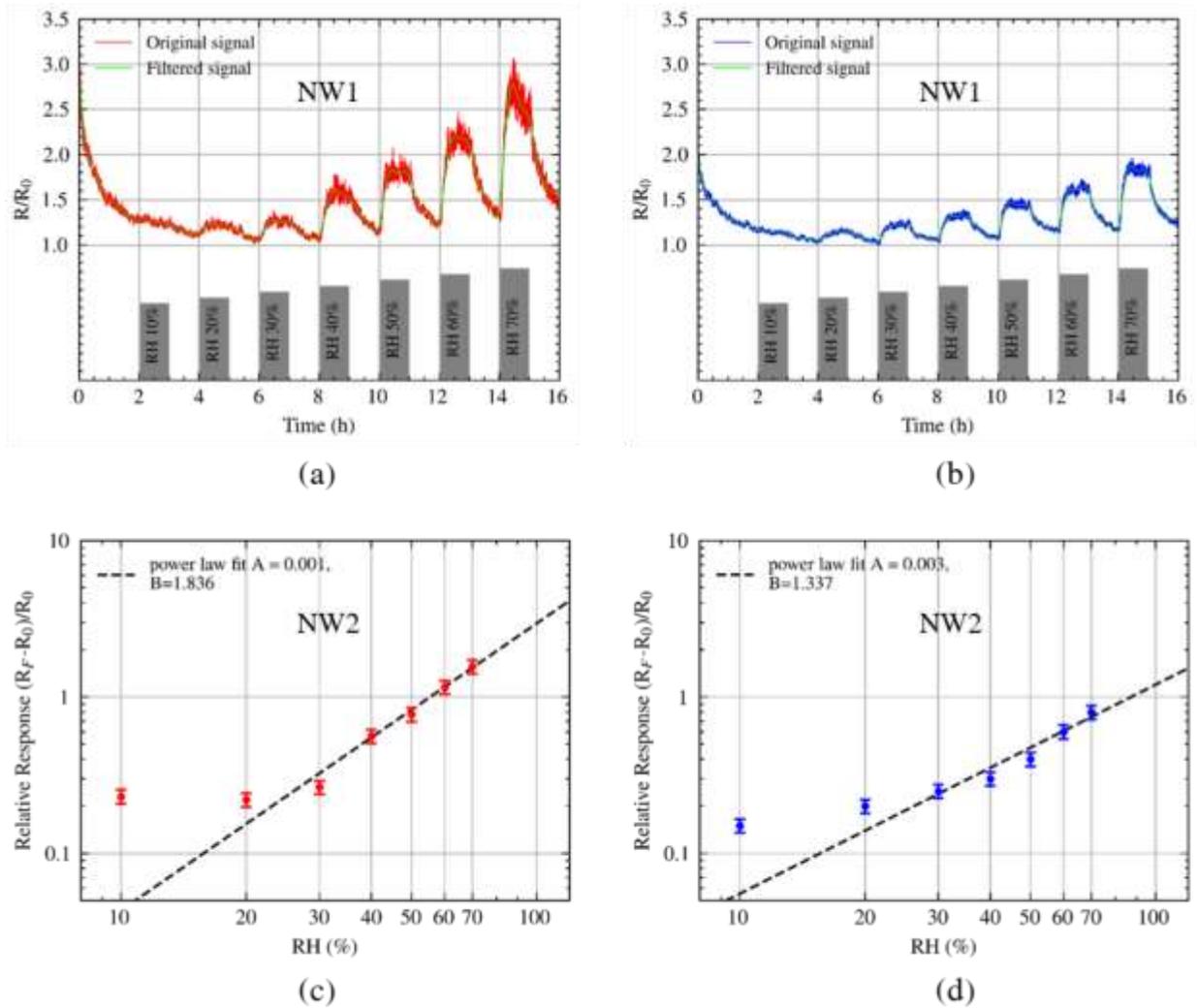

Figure 2: Dynamic variation of the NW twin resistance normalized to the initial (0% RH) value, towards changes in RH from 10% to 70%. Variations of relative resistance are reported for *NW* 1 (a) and *NW* 2 (b), together with the corresponding calibration curves for *NW* 1 (c) and *NW* 2 (d). Error bars indicate standard deviations. For both signals, a 5th-order digital Butterworth filter was applied with a cutoff frequency of 2.5 mHz; the filtered results are reported as thin green lines. This filter attenuates fluctuations in the signals with a period shorter than 400 seconds.

During each sensing cycle, we observed an increase in device resistance upon exposure to humidity, followed by a subsequent drop in resistance to the baseline value during the recovery period (dry air flushing), with the maximum resistance value increasing with the RH value for each cycle. The NWs were sensitive to RH in concentrations down to 20%. Prior to reaching a relative humidity (RH) of 30%, the NW twins exhibited nearly identical resistance trend. However, as the RH exceeded 40%, a deviation was observed. For the sake of comparison, we define relative response (RR) as $(R-R_0)/R_0$, where R is the steady state resistance in gas and $R_0$ is the steady state resistance in air. Specifically, the relative response of *NW* 1 began to amplify compared to that of *NW* 2. At an RH of 70%, the differential in relative response reaches its maximum value (approximately 1.4-fold). The electrical noise associated with the *NW* 1 also increases at higher RH. The noise analysis will be tackled in the discussion section.

**$NO_2$ detection**

In Figure 3, we present the results of the sensing tests of the twin NW devices against $NO_2$ for concentrations ranging from 2 to 9 ppm in 1 ppm increments.

To perform the $NO_2$ sensing tests, we employed a testing cycle that comprised of a 2-hour exposure period during which the sensors were exposed to a input air flow with specific and constant $NO_2$ concentration, followed by a 2-hour recovery period during which the samples were allowed to recover to their baseline resistance in dry air. As expected, NW twin devices exhibited sensitivity to $NO_2$ even at the lowest tested concentrations. However, we observed that *NW* 1 displayed a higher response to $NO_2$ than *NW* 2 when concentrations exceeded 4 ppm. Furthermore, both *NW* 1 and *NW* 2 exhibited a saturation behavior at concentrations higher than 5 ppm.

During each sensory cycle, we noted a transient increase in the device resistance when subjected to humidity and $NO_2$. This increase was followed by a decrease towards baseline resistance levels during the recovery phase, which involved dry air flushing. Notably, the InAs NW twins were able to desorb $NO_2$ molecules solely through air flushing. This does not occur with NWs with different compositions, such as metal oxide NWs, that necessitate high operating temperatures or UV desorption at ambient conditions for the removal of gas molecules from their surfaces.[23] Remarkably, slight differences occurred in the responses of the InAs NW twin devices, and a noise analysis of the experimental datasets was carried out in order to elucidate the nature and the sources of such differences.

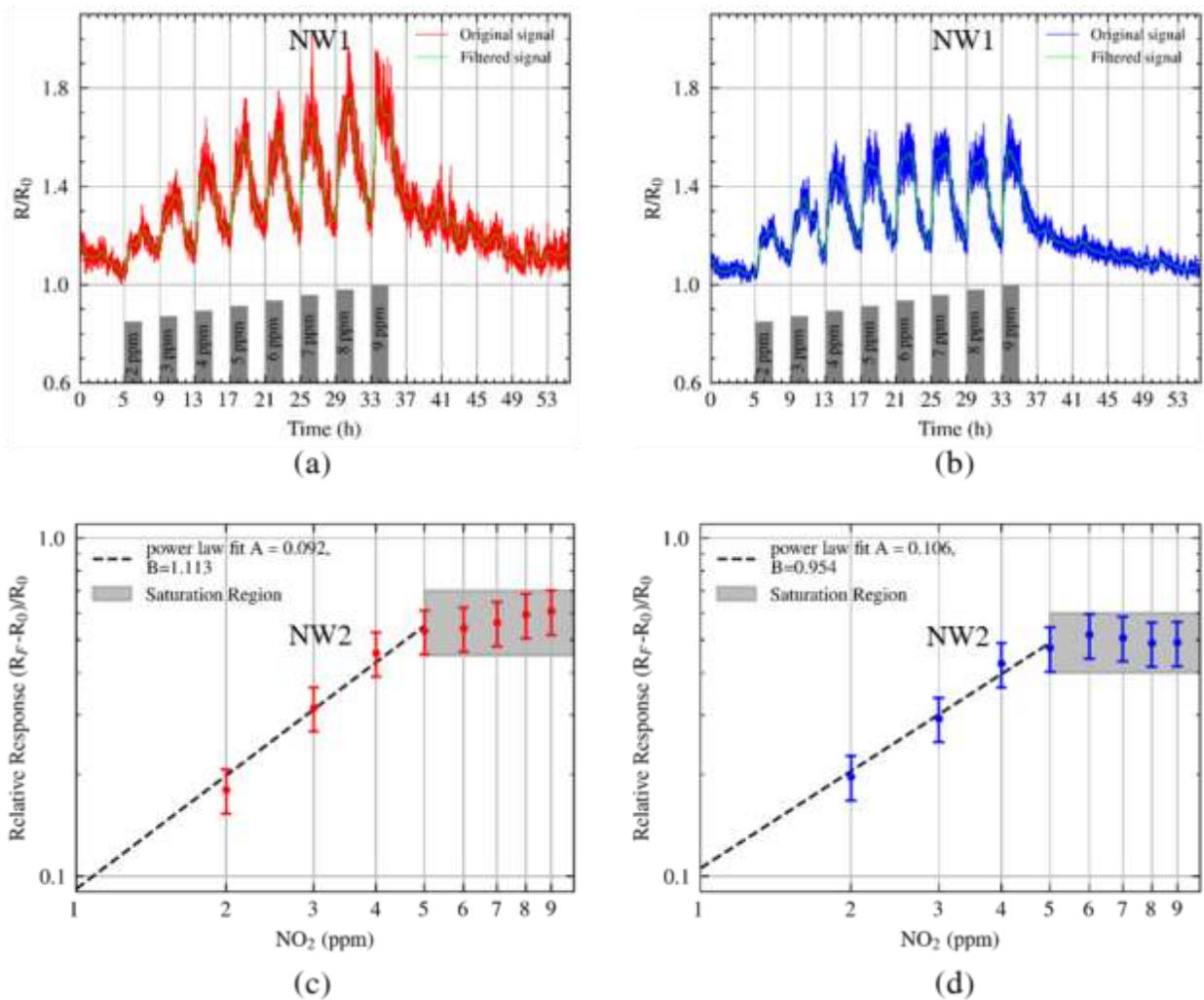

Figure 3: Dynamic response of the NW twins versus changes in NO$_2$ from 2 ppm to 9 ppm. Variations of relative resistance are reported for *NW* 1 (a) and *NW* 2 (b), along with their corresponding calibration curves for *NW* 1 (c) and *NW* 2 (d). For both signals, a 5th-order digital Butterworth filter was applied with a cutoff frequency of 1/900 Hz, and the results are shown as thin green lines in panels a and b. This filter attenuates fluctuations in the signals with a period shorter than 900 seconds.

## Signal Fluctuations Analyses

In each measurement, the NW twin devices generate a pair of voltage signals that, in the ideal case of identical devices, should also be identical. While from a bird-eye view similar trends can be identified in the signals measured in the two nanowires, however, at a closer look each pair of measured signals display clear differences between the two individual voltage traces of each NW, both in terms of signal intensity and signal-to-noise ratio. In order to quantitatively address the similarity of different series of datasets measured simultaneously in NW twin devices, we applied the correlation analysis method and calculated auto-correlation (AC) and cross-correlation (CC) functions for all the experimental signals pairs. The goal was understanding the source of the signal fluctuations: to this aim, we carried out a careful analysis of the noise and of its time dependence, both for individual NWs and by checking the temporal correlations among the measurement fluctuations in the NW twin devices.

For this analysis we considered the extended set of non-normalized data shown in Figure 4, reported as resistance (R, panel (a)) or conductance (S, panel (b)) and covering the full measurement time interval, which exceeds fifty hours. In both panels, orange lines repre- sent bi-exponential fits calculated for the different chamber filling/emptying periods. These datasets include those already reported, as renormalized resistance, in panels (a) and (b) of the Figures 2 and 3.

It can observed that, the higher the resistance $R$ (i.e. the measured voltage with nominally constant current), the higher the noise. Moreover, higher resistances are measured at higher concentrations of contaminants. By considering instead the conductance $S = \frac{1}{R}$ of the NWs as a function of time (Figure 4(b)), the noise seems more constant. This behavior is quantitatively analyzed in panels (c) and (d) of Figure 4, showing the standard deviations of the signals as a function of the signal intensities. To generate these graphs, the time dependence of $R$ and $S$ was fitted, within each chamber filling/emptying period of time, with a biexponential-decay-with-offset function (orange lines in Figure 4(a,b)), and each point in panels (c) and (d) represents the quadratic average of the fit residuals, within a window of 40 points (corresponding to a 3.3 minutes time window) centered every 20 points within each chamber filling/emptying periods. As expected, while the standard deviation $\sigma_R$ exhibits a a strong dependence on $R$, the dependence of $\sigma_S$ upon $S$ is definitely less pronounced. Besides, with the lowest amounts of RH and $NO_2$, signals and their standard deviations tends to similar values for each NW (lowest $S$ and highest $R$ for each combination of NW and contaminant). At such low amounts of the analytes, $\sigma_S$ is similar for the two NWs, just slightly higher for NW2. The signal-to-noise ratio (SNR) is larger for NW2, which has a smaller starting resistance (higher starting conductivity) than NW1. In all cases, with the increase of the contaminant, the SNR decreases. Finally, the dependence of $\sigma_S$ on the amount of RH appears quite negligible for both NWs. Notably, at high values of $NO_2$ the noise is bigger than at high values of RH, for corresponding values of $S$ and $R$: this indicates that (i) the noise not only depends on the measurement setup, but is also affected by the interactions of the contaminants with the NW surfaces, and (ii) these interactions are less stable (or cause bigger signal deviations) in the case of $NO_2$.

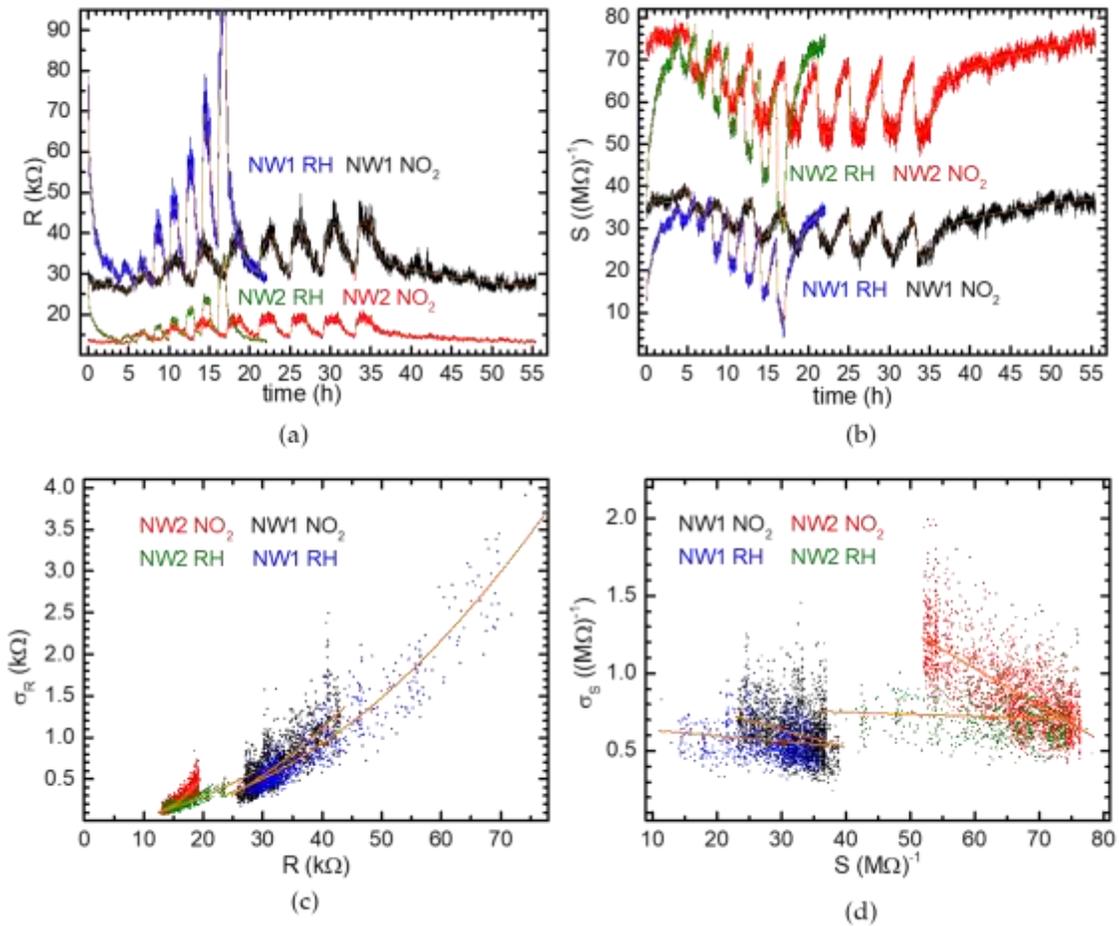

Figure 4: (a) Electrical resistance, R, and (b) conductance, S, as a function of time for NW twin devices exposed to RH and NO$_2$. The data are the same from which the ones reported in Fig. 4 (a) and (b) and Fig. 5 (a) and (b); refer to those figures for the RH and the concentration of NO$_2$. Orange lines are bi-exponential fits. For the composition of the gas flown at different times see Figures 2 and 3; there is an additional filling phase at RH of 80% between 16 and 17 h in the RH case, but those data have not been used in any reported analysis. (c) Standard deviation of resistance and (d) conductance, extracted from the fit residuals, as function of the fitted value of R and S, respectively. Orange lines are parabolic fits in (c) and linear fits in (d).

Signal fluctuations can stem from two phenomena: the change in the local concentration of the analyte, and intrinsic dynamics in the sensor behavior. The latter may trivially account for the device response time, or being related to the occurrence of metastable configurations in the NW-gas molecule system, changing the electrostatic environment surrounding the NW surface. The two NWs are so close to each other that they can virtually be considered as occupying the same location on the fabrication substrate. Therefore, any change in the local gas composition or concentration should impact in the same way on the electrical transport in both NWs. Consistently, the main electrical transport features measured in both NWs display very similar overall time dependence. In more detail, during the chamber filling step, signal variation occurs with a time constant around 0.1h (six minutes) for both NWs in the case of RH, followed by a slower drift, especially for NW1. For $NO_2$, the time constant increases to a value between 0.13h and 0.26h (8 and 16 minutes) for the higher concentrations, or seemingly up to 40 minutes for the lowest tested $NO_2$ concentration (NW2 exhibits the slowest response). In all cases, the measured time constants are consistent with the time interval (approximately 10 minutes) necessary for changing the gas inside the measurement chamber (volume around one liter, used gas flow of 0.1 l/min at standard temperature and pressure). Instead, during the chamber emptying steps, the observed signal dynamics was overall significantly slower. In particular, in the initial phase of the emptying process of the chamber, monoexponential fits return time constants between 0.3 and 0.8 hours for RH, and around 1 h for $NO_2$. In the longer emptying phase, the fits return even longer time constants: about 7 h for $NO_2$ for both NWs, and about 2 (1.4) hours for NW1 (NW2) exposed to RH. Tentatively, such slow dynamics could be ascribed to the desorption of gas molecules previously trapped at the NW surfaces, or to the degassing from the inner surface of the chamber.

The amplitudes of the observed fluctuations in $S$ are relatively similar in all the investigated combinations of NWs and gas species. For this reason, we focused on the electrical conductivity experimental signal datasets for performing the analysis yielding to the estimation of auto-correlation and cross-correlation functions of the measured signals. These allow to visualize any fast time dependence occurring in a signal by addressing the characteristic times of the fluctuations, and are calculated according to the following expression:

$$G_{ab}(\tau) = \overline{S_a(t) - \overline{S_a(t)}\ \ S_b(t+\tau) - \overline{S_b(t+\tau)}}\ ,$$

where the subscripts *a* and *b* label NW1 and/or NW2, the angular bracket indicate average over time *t*, $S_{a,b}$ are the two signals and $G_{ab}$ represents the cross-correlation (CC) between $S_a$ and $S_b$, or the auto-correlation (AC) function $G_{aa}$ if *a* is equal to *b*. $\tau$ indicates the time lag (or interval) and represents the independent variable. For a stationary process, $\overline{S_a}$ is taken as the time-independent average of $S_a$. In our case, in order to correct the impact of the very slow variations, we assume $\overline{S_a}$ as given by the fitted curves shown in Figure 4(b). In case of environmental fluctuations, the two NWs composing the NW twin devices are close enough to experience exactly the same fluctuations. Assuming that the two twin sensors are nominally identical, the conductance fluctuations observed in each nanodevice should be identical as well. In this framework, the occurrence of a specific feature in the curve $G_{aa}$ as a function of lag time but not in the $G_{ab}(\tau)$ curve indicates that signal fluctuations are not of environmental origin, but instead arise from a cause that is local to each NW (e.g., minor morphological differences between the NW twin devices).

Figure 5 reports the AC and CC curves calculated for the two single NW-based devices labeled NW1 and NW2 operating under exposure to RH and $NO_2$ gases. Both AC and CC curves have been calculated within each chamber filling/emptying time intervals (or some fractions), and then averaged. Figure 5 (a) and (b) report the CC curves together with the AC ones calculated for the entire measurement times, displaying the same datasets with a linear (5 (a)) or a logarithmic (5 (b)) scale for the independent variable, the lag time $\tau$ in hours. Notably, for low — $\tau$ —, the CC curves for both gases display significant lower values with respect to the corresponding AC ones. This indicates that the fluctuations experi- mentally observed in the conductance of the two devices very likely do not originate from fluctuations in the environment, but rather from device-specific features. The autocorrela- tion curves for $NO_2$ are stronger and longer-lasting with respect to their counterparts for RH. Besides, the AC curves extracted for NW1 and NW2 decay with very similar trend in experiments with variation of both RH and $NO_2$. Exponential fits return for RH an average decay time in the range 0.022-0.027 h (1-2 minutes), and for $NO_2$ an average decay times about 0.085 h (5 minutes). In a biexponential fit, the two components read 0.004-0.005h (15-20 sec) and 0.035-0.055 h (2-3 minutes) for RH, while for $NO_2$ they read 0.009-0.012 h (0.5-0.75 minutes) and 0.1-0.14 h (6-8.5 minutes).

If the conductance fluctuations depend on the presence of the gas molecules, then they might depend on the gas concentration. To clarify these aspects, we compared the AC curves of both NWs for two different concentrations of $NO_2$ (Figure 5 c)) and RH (Figure 5 d)). In panel 5 c), the label "low $NO_2$" refers to AC functions calculated starting from the signal region corresponding to the last longer emptying phase (time longer than 37.27 hours) corresponding to an estimated concentration of 4 ppm, while the label "high $NO_2$" refers to a signal region at or close to the plateau characteristic of the filling phases, corresponding to 9 ppm; solid lines correspond to the curves averaged on the whole experiment reported also in panel 7(b) and are shown here for comparison. In panel 5 d), the label "low RH" refers to times above 17.79 h corresponding to 30 % RH, while "high RH" refers to a signal region at or close to the plateau during the filling phases, corresponding to 70 % RH; solid lines correspond to the curves averaged on the whole experiment reported also in panel 7(b), and are shown here for comparison. Interestingly, the AC curves for $NO_2$ display slower decay with respect to RH, even in the low-contaminant case. This implies that the explored $NO_2$ concentration was never low enough to have negligible impact on the noise, making it very difficult to discriminate between two different noise sources, namely, electrical measurements and fluctuations in the interactions between NWs and gas molecules.

Finally, we notice a few slow macroscopic signal fluctuations occurring especially in the tail of the last $NO_2$ emptying phase (more evident for NW1; see Figure 4 d). Similar slow fluctuations can be observed in all fit residuals for the last emptying phase, with typical duration of about one hour for $NO_2$ and half an hour for RH, i.e., close to the times charac- teristic of signal changes in the emptying phases. Such fluctuations occur for both NWs and are uncorrelated, revealing that their origin cannot be ascribed to differences in the environ- ment of the twin NWs. Moreover, the occurrence of these fluctuations might play a role in the onset of the non-monotonic behavior and the negative values observed in the correlation curves at the latest lag times, as visible in Figure 5. Keeping in mind the fluctuation-dissipation theorem, we tentatively rationalize these experimental outcomes as follow. The slow dynamics in the emptying phases is likely ascribable to a slow response of the NWs upon removal of the gas molecules, most probably because of the slow desorption of the gas molecules from the NW surfaces, especially in the NW regions close to the ohmic contacts. This scenario indicates a possible route for enhancing the response time of InAs NW-based gas sensors, namely, increasing the operating temperature of the NW, specially when the gas concentration is relatively low.

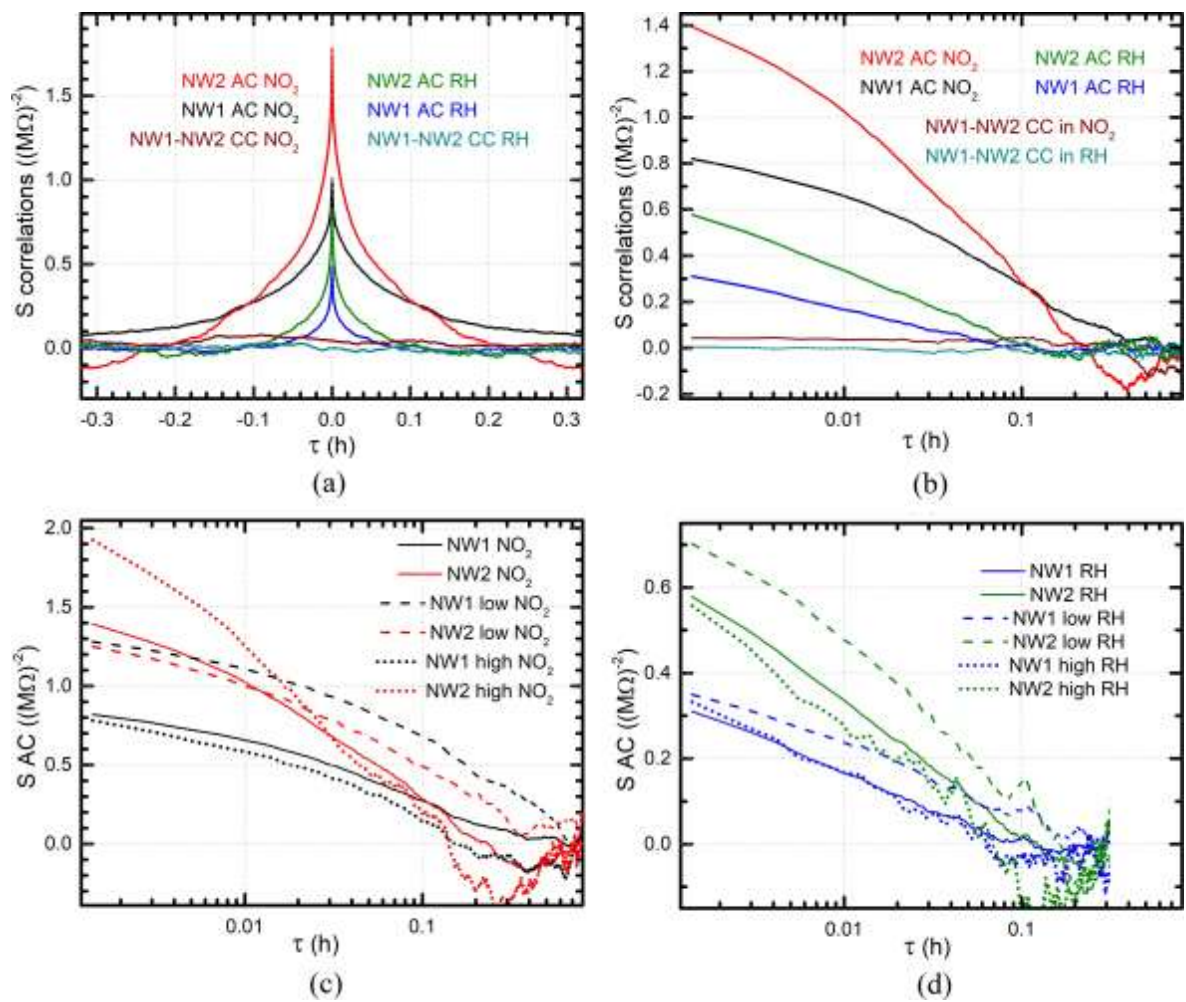

Figure 5: AC and CC curves for an InAs NW-based twin sensor and two different gases: (a) linear plot and (b) semilogaritmic plot. c) AC curves for the two NW devices operating under NO$_2$ exposure; "high NO$_2$" labels AC curves accounting for from multiple time intervals (13.5h-15h, 17.5h-19h, 21.5h-23h, 25.5h-27h, 29.5h-31h and 33.5h-35h; "low NO$_2$" times longer than 37.27 h. d) AC curves for the two NW devices operating under humidity exposure; high RH indicates curves calculated from the conductance in the time intervals (6.55h-7.03h, 8.56h-9.03h, 10.57h-11.04h, 12.58h-13.05h, 14.58h-15.06h); "low NH" accounts for times longer than 17.79 h.

## Conclusions

We reported the combined experimental measurement and signal correlation analysis of the electrical response of InAs nanowire-based twin devices, used to probe different gases in synthetic air. We engineered pairs of nanowire-based nominally identical sensors (twin sensors) and simultaneously measured their response upon exposure to relative humidity or $NO_2$ flux, by resorting on a custom readout interface designed and realized in-house. Auto- correlation and cross-correlation analysis were performed on simultaneously measured pairs of current signals, suggesting changes at nanoscopic levels (i.e., local to each nanodevice) - related to the nanostructure and device morphology - as the major source of lack of cross- correlation in the InAs nanowire-based twin devices, and ruling out a prominent role of inhomogeneities in the chemical environment.

# Experimental Section

## Nanowire growth and device nanofabrication

Wurtzite InAs NWs were grown by gold-assisted Chemical Beam Epitaxy (CBE) on InAs (111)B substrates, using trimethylindium (TMIn) and tertiarybutylarsine (TBAs) metallorganic precursors.[29] Gold nanoparticle catalysts were obtained by dewetting (at 540 ± 10 °C under TBA flow for 20 min) of a 0.5nm thick gold film, previously evaporated on the InAs substrates. The NWs were grown at a temperature of 465±10 °C, with TMIn and TBA line pressures of ≈0.9 and ≈0.3 Torr, respectively, for a time of ≈45 min. Ditertiarybutylselenide (DtBSe) with a line pressure of 0.10 Torr was used as n-doping source . After the growth, the NWs were mechanically detached from the growth substrate and dispersed in isopropyl alcohol (IPA) by sonicating the substrate in IPA. A droplet of the IPA/NWs solution was then deposited by dropcasting onto a $SiO_2/SiO_2$++ substrate pre-patterned with markers for nanofabrication). The contact electrodes were patterned by a single step aligned electron beam lithography (EBL). After the development, and immediately before the evaporation of a bilayer metal contact (Cr/Au, 10/100 nm), the NW contact areas were exposed to an ammonium polysulfide $(NH_4)_2$S-based solution to promote the formation of low-resistance ohmic contacts. The chip was then attached to a standard Dual-in-Line chip carrier using a conductive silver paste and the devices were wire bonded to the carrier.

## Customized electronic circuitry for simultaneous sensing with two nanowires

In order to simultaneously measure the electrical response of each NW, we designed a dedicated electronic circuitry to feed a constant current through the two NWs, and we measured the voltage drop across the NWs: this allows us to avoid heating-induced changes in the NW response. In fact, self-heating in nanostructures, including NWs, can occur even at very low electrical power.[30] While this effect is less pronounced in substrate-supported NWs with respect to suspended ones,[31] it must still be considered. To avoid NW damage due to self-heating, a stable current of 2 µA was feed across the nanostructures during the experiments. The circuit also allows for measuring the resistance of the NW twin devices and comprises six key components: two constant current source circuits based on LM334,[32] a voltage amplifier circuit in the non-inverting operational amplifier configuration (Figure 6 b) based on AD8656,[33] an analog-to-digital converter (ADC) based on ADS1220,[34] an Arduino UNO (MCU) board [35] (Figure 6 a), an ambient temperature sensor SHT33-dis to measure the temperature outside the test chamber [36], and a PC. In the circuit (Figure 6 b), the resistor element labeled "NW" corresponds to the device.

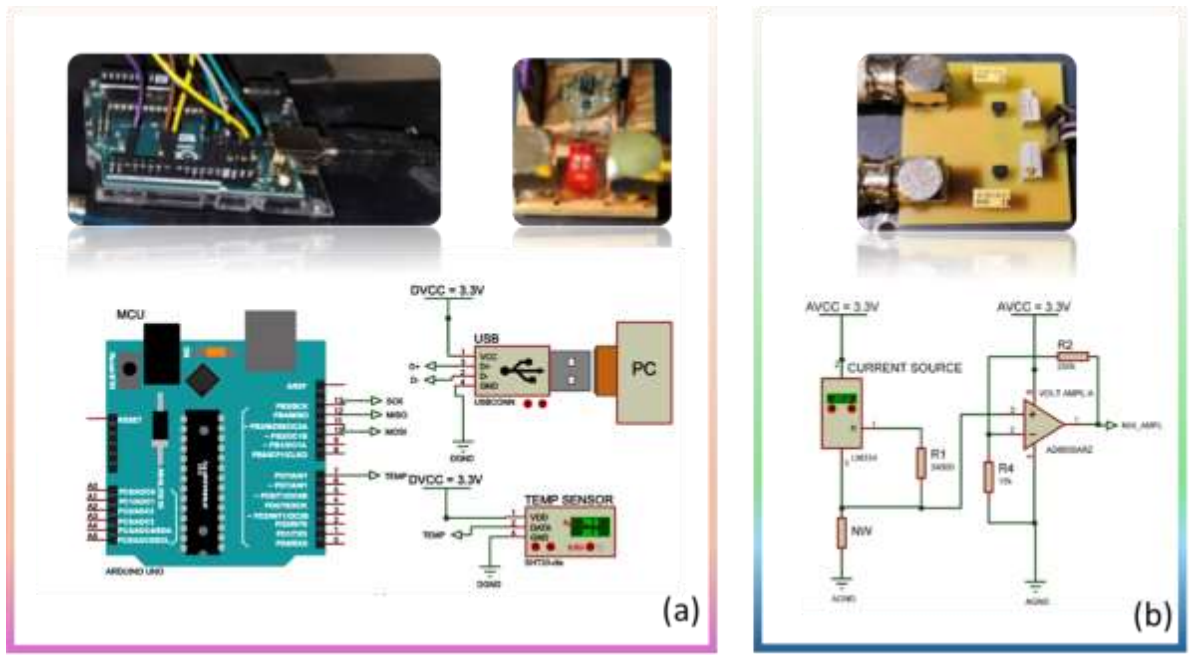

Figure 6: Schematic block diagram and images of (a) data processing and data transmission; (b) NWs resistance voltage conversion circuit (one for each nanowire).

As shown in Figure 6, the current sources are configured to provide a consistent current to the NWs, set at 2µA, determined by the resistor R1.[32] As the current flows through the NWs, it generates a voltage drop according to Ohm's law. The voltage signal of each NW is then amplified by the voltage amplifier, with a gain factor of 16.5. Consequently, when the resistance of the NWs reaches its peak value of 100 kOhm, the resultant voltage is approximately 3.5 V. This value represents the maximum input voltage for the ADC, ensuring optimal resolution for current measurements. The ADC transmits information regarding the voltage drop observed across the NWs to the MCU via a serial peripheral interface. The MCU then processes these data. To account for the temperature drift of the current source, a temperature compensation mechanism was integrated into the circuit design. This mechanism comprises a temperature sensor that is capable of measuring the temperature around the current source circuit and transmitting this information to the MCU for processing. By taking into account the data obtained from the ADC and the temperature sensor, the MCU computes the true value of the resistance of the NWs, thereby correcting for any temperature-induced variations in the current source output. In real-time, the PC retrieves data on the resistance of the NWs via a USB interface connected to the MCU. This information is then visualized and concurrently saved into a text files, facilitating subsequent analysis.

**Gas sensors testing setup**

Sensing tests were carried out in a stainless steel test chamber (1000 cm$^3$) with a constant flux of synthetic air (100 cm$^3$/min). The ratio between gas flux and test chamber size was chosen to avoid turbulence inside the chamber, at the expense of the chamber filling time. $NO_2$ at the ppm concentration diluted in dry air is supplied from certified bottles, while relative humidity is obtained by properly diluting air carrier saturated with humidity.[35]

**Acknowledgements**

CB and EM acknowledges the support from Antares Vision; CB and GF acknowl- edges the support from European Union-Next Generation EU M4C2 1.1 under Grant PRIN 2022JZAA9W–SENSEPLANET. LR and FR acknowledges the support from the National Recovery and Resilience Plan (PNRR), Mission 04 Component 2 Investment 1.5 – NextGen- erationEU, Call for tender n. 3277 dated December 30, 2021 (Award Number: 0001052 dated June 23, 2022). VD acknowledges the support from the Project "Network 4 Energy Sustainable Transition—NEST", Spoke 1, Project code PE0000021, funded by the European Union—NextGenerationEU under the National Recovery and Resilience Plan (NRRP), Mission 4, Component 2, Investment 1.3, Call for tender No. 1561 of 11.10.2022 from italian MUR.

# References

(1) Milone, A.; Monteduro, A. G.; Rizzato, S.; Leo, A.; Di Natale, C.; Kim, S. S.; Maruccio, G. Advances in Materials and Technologies for Gas Sensing from Environmental and Food Monitoring to Breath Analysis. *Advanced Sustainable Systems* **2022**, *7*.

(2) Patolsky, F.; Lieber, C. M. Nanowire nanosensors. *Materials Today* **2005**, *8*, 20–28.

(3) He, B.; Morrow, T. J.; Keating, C. D. Nanowire sensors for multiplexed detection of biomolecules. *Current Opinion in Chemical Biology* **2008**, *12*, 522–528.

(4) Rocci, M.; Demontis, V.; Prete, D.; Ercolani, D.; Sorba, L.; Beltram, F.; Pennelli, G.; Roddaro, S.; Rossella, F. Suspended InAs Nanowire-Based Devices for Thermal Conductivity Measurement Using the 3 Method. *Journal of Materials Engineering and Performance* **2018**, *27*, 6299–6305.

(5) Demontis, V.; Zannier, V.; Sorba, L.; Rossella, F. Surface Nano-Patterning for the Bottom-Up Growth of III-V Semiconductor Nanowire Ordered Arrays. *Nanomaterials* **2021**, *11*, 2079.

(6) Demontis, V.; Rocci, M.; Donarelli, M.; Maiti, R.; Zannier, V.; Beltram, F.; Sorba, L.; Roddaro, S.; Rossella, F.; Baratto, C. Conductometric Sensing with Individual InAs Nanowires. *Sensors* **2019**, *19*, 2994.

(7) Zagaglia, L.; Demontis, V.; Rossella, F.; Floris, F. Semiconductor nanowire arrays for optical sensing: a numerical insight on the impact of array periodicity and density. *Nanotechnology* **2021**, *32*, 335502.

(8) Zagaglia, L.; Demontis, V.; Rossella, F.; Floris, F. Particle swarm optimization of GaAs-AlGaAS nanowire photonic crystals as two-dimensional diffraction gratings for light trapping. *Nano Express* **2022**, *3*, 021001.

(9) Floris, F.; Fornasari, L.; Marini, A.; Bellani, V.; Banfi, F.; Roddaro, S.; Ercolani, D.; Rocci, M.; Beltram, F.; Cecchini, M.; Sorba, L.; Rossella, F. Self-Assembled InAs Nanowires as Optical Reflectors. *Nanomaterials* **2017**, *7*, 400.

(10) Bârsan, N. Transduction in Semiconducting Metal Oxide Based Gas Sensors - Implications of the Conduction Mechanism. *Procedia Engineering* **2011**, *25*, 100–103.

(11) Zhang, Y.; Kolmakov, A.; Chretien, S.; Metiu, H.; Moskovits, M. Control of Catalytic Reactions at the Surface of a Metal Oxide Nanowire by Manipulating Electron Density Inside It. *Nano Letters* **2004**, *4*, 403–407.

(12) Mirzaei, A.; Lee, J.-H.; Majhi, S. M.; Weber, M.; Bechelany, M.; Kim, H. W.; Kim, S. S. Resistive gas sensors based on metal-oxide nanowires. *Journal of Applied Physics* **2019**, *126*.

(13) PMirzaei, A.; Leonardi, S. G.; Neri, G. Detection of hazardous volatile organic compounds ( VOCs ) by metal oxide nanostructures-based gas sensors : A review. *Ceramics International* **2016**, *42*.

(14) Baratto, C.; Golovanova, V.; Faglia, G.; Hakola, H.; Niemi, T.; Tkachenko, N.; Nazarchurk, B.; Golovanov, V. On the alignment of ZnO nanowires by Langmuir – Blodgett technique for sensing application. *Applied Surface Science* **2020**, *528*, 146959.

(15) Comini, E.; Baratto, C.; Faglia, G.; Ferroni, M.; Vomiero, A.; Sberveglieri, G. Quasi-one dimensional metal oxide semiconductors: Preparation, characterization and application as chemical sensors. *Progress in Materials Science* **2009**, *54*, 1–67.

(16) Tan, H. M.; Hung, C. M.; Ngoc, T. M.; Nguyen, H.; Hoa, N. D.; Duy, N. V.; Hieu, N. V. Novel Self-Heated Gas Sensors Using on-Chip Networked Nanowires with Ultralow Power Consumption. *Applied Materials and Interfaces* **2017**, *7*.

(17) Baratto, C. Growth and properties of ZnO nanorods by RF-sputtering for detection of toxic gases. *RSC Advances* **2018**, *8*, 32038–32043.

(18) Kolmakov, A. Some recent trends in the fabrication, functionalisation and characterisation of metal oxide nanowire gas sensors. *International Journal of Nanotechnology* **2008**, *5*, 450.

(19) Hernández-Ramírez, F.; Tarancón, A.; Casals, O.; Arbiol, J.; Romano-Rodríguez, A.; Morante, J. R. High response and stability in CO and humidity measures using a single $SnO_2$ nanowire. *Sensors and Actuators B: Chemical* **2007**, *121*, 3–17.

(20) Meng, G.; Zhuge, F.; Nagashima, K.; Nakao, A.; Kanai, M.; He, Y.; Boudot, M.; Takahashi, T.; Uchida, K.; Yanagida, T. Nanoscale Thermal Management of Single $SnO_2$ Nanowire: pico-Joule Energy Consumed Molecule Sensor. *ACS Sensors* **2016**, *1*, 997–1002.

(21) Donarelli, M.; Ferroni, M.; Ponzoni, A.; Rigoni, F.; Zappa, D.; Baratto, C.; Faglia, G.; Comini, E.; Sberveglieri, G. Single Metal Oxide Nanowire devices for Ammonia and Other Gases Detection in Humid Atmosphere. *Procedia Engineering* **2016**, *168*, 1052–1055.

(22) Baratto, C.; Kumar, R.; Faglia, G.; Vojisavljević, K.; Malič, B. p-Type copper aluminum oxide thin films for gas-sensing applications. *Sensors and Actuators B: Chemical* **2015**, *209*, 287–296.

(23) Zhang, D.; Liu, Z.; Li, C.; Tang, T.; Liu, X.; Han, S.; Lei, B.; Zhou, C. Detection of NO$_2$ down to ppb Levels Using Individual and Multiple In$_2$O$_3$ Nanowire Devices. *Nano Letters* **2004**, *4*, 1919–1924.

(24) Du, J.; Liang, D.; Tang, H.; Gao, X. P. InAs Nanowire Transistors as Gas Sensor and the Response Mechanism. *Nano Letters* **2009**, *9*, 4348–4351.

(25) Lynall, D.; Tseng, A. C.; Nair, S. V.; Savelyev, I. G.; Blumin, M.; Wang, S.; Wang, Z. M.; Ruda, H. E. Nonlinear Chemical Sensitivity Enhancement of Nanowires in the Ultralow Concentration Regime. *ACS Nano* **2020**, *14*, 964–973.

(26) Offermans, P.; Crego-Calama, M.; Brongersma, S. H. Gas Detection with Vertical InAs Nanowire Arrays. *Nano Letters* **2010**, *10*, 2412–2415.

(27) Zhang, X.; Fu, M.; Li, X.; Shi, T.; Ning, Z.; Wang, X.; Yang, T.; Chen, Q. Study on the response of InAs nanowire transistors to H2O and NO2. *Sensors and Actuators B: Chemical* **2015**, *209*, 456–461.

(28) Ullah, A. R.; Joyce, H. J.; Tan, H. H.; Jagadish, C.; Micolich, A. P. The influence of atmosphere on the performance of pure-phase WZ and ZB InAs nanowire transistors. *Nanotechnology* **2017**, *28*, 454001.

(29) Prete, D.; Dimaggio, E.; Demontis, V.; Zannier, V.; Rodriguez-Douton, M. J.; Guazzelli, L.; Beltram, F.; Sorba, L.; Pennelli, G.; Rossella, F. Electrostatic Control of the Thermoelectric Figure of Merit in Ion-Gated Nanotransistors. *Advanced Functional Materials* **2021**, *31* .

(30) Prades, J. D.; Jimenez-Diaz, R.; Hernandez-Ramirez, F.; Barth, S.; Cirera, A.; Romano-Rodriguez, A.; Mathur, S.; Morante, J. R. Ultralow power consumption gas sensors based on self-heated individual nanowires. *Applied Physics Letters* **2008**, *93* .

(31) Chikkadi, K.; Muoth, M.; Maiwald, V.; Roman, C.; Hierold, C. Ultra-low power operation of self-heated, suspended carbon nanotube gas sensors. *Applied Physics Letters* **2013**, *103* .

(32) Available online: https://www.ti.com/lit/ds/symlink/lm334.pdf (accessed on 17 October 2023).

(33) Available online: https://www.analog.com/media/en/technical-documentation/ datasheets/AD8655_8656.pdf (accessed on 17 October 2023).

(34) Available online: https://www.ti.com/lit/ds/symlink/ads1220.pdf (accessed on 17 October 2023).

(35) Endres, H.-E.; Jander, H. D.; Göttler, W. A test system for gas sensors. *Sensors and Actuators B: Chemical* **1995**, *23*, 163–172.